\documentstyle[epsfig]{mn}

\def\Msun{\ifmmode{~{\rm M}_\odot}\else${\rm M}_\odot$~\fi}
\def\kms{\ifmmode{$~km\thinspace s$^{-1}}\else km\thinspace s$^{-1}$\fi}
\def\veff{v_{\mathrm eff}}
\def\fwd{F_{\mathrm WD}}
\def\etal{et al~}
\def\ee{\end{equation}}
\def\be{\begin{equation}}
\def\masyr{{\rm mas}\, {\rm yr}^{-1}}

\title{No nearby counterparts to the moving objects in the Hubble Deep Field}

\author[Chris Flynn, J.~Sommer-Larsen, Burkhard Fuchs, David S. Graff, Samir
Salim] {Chris Flynn$^1$, J.~Sommer-Larsen$^2$, B.~Fuchs$^3$, David
S. Graff$^4$ \& Samir Salim$^4$\\ $^1$Tuorla Observatory, Piikki\"o, FIN-21500,
Finland\\ $^2$Theoretical Astrophysics Center, Juliane Maries Vej 30,
Copenhagen, Denmark\\ $^3$Astronomisches Rechen-Institut, M\"onchhofstrasse
12-14, Heidelberg, Germany\\ $^4$The Ohio State University Depts. of Physics
and Astronomy, Columbus, OH 43210, USA}

\begin{document}

\maketitle

\voffset=-1.0cm     

\begin{abstract} 

  Ibata et al (1999) have recently discovered faint, moving objects in the
Hubble Deep Field. The quantity, magnitudes and proper motions of these objects
are consistent with old white dwarfs making up the Galactic dark halo. We
review a number of ground-based proper motion surveys in which nearby dark halo
white dwarfs might be present, if they have the colours and absolute magnitudes
proposed. No such objects have been found, whereas we argue here that several
times more would be expected than in the Hubble Deep Field.  We conclude it is
unlikely that hydrogen atmosphere white dwarfs make up a significant fraction
of the halo dark matter. No limits can be placed yet on helium atmosphere
dwarfs from optical searches.

\end{abstract}


\section{Introduction}

  Ibata et al (1999, hereafter IRGS99) have detected faint moving objects in the
Hubble Deep Field (HDF).  These objects have proper motions, apparent
magnitudes and colours which are consistent with a Galactic dark halo made of
old white dwarfs (WDs). 

  There are however reasons to suppose that white dwarfs make up only a limited
amount of dark matter.  They would over-pollute the Universe with carbon by a
factor of 100 (Fields, Freese \& Graff 1998).  Although Chabrier (1999) has
suggested that the above result may be model dependant, robust limits on the
cosmic density of white dwarfs can be placed using helium and deuterium
abundances (Fields, Freese \& Graff 1999) and limits on background infrared
photon number density (Graff, Freese, Walker \& Pinsonneault 1999).  These
limits caused Hansen (1999a) to postulate the existence of beige dwarfs,
degenerate massive objects that are not stellar remnants, and would escape the
above restrictions.  None of these objections are so robust that ways of
circumventing them cannot be found.  Some of the standard objections to white
dwarfs as the halo dark matter are discussed by Richer (1999).

  If the IRGS99 objects are halo white dwarfs, several nearby counterparts to
such objects might be located in ground-based, wide area, proper motion
surveys, in which they would appear as faint, high proper motion objects.
Several such white dwarfs may have been found already. Hodgkin et al (1999)
have followed up the Hambly et al (1999) discovery of a very cool, halo white
dwarf candidate. The object is at a distance of $28\pm4$ pc and has velocity
components consistent with belonging to a halo population. Harris et al (1999)
have also located a very, low luminosity cool, white dwarf, although it's
population type (disk or halo) is not yet known.

  Such objects could be related to the objects found by IRGS99.  In this paper,
we examine the IRGS99 proposed population of dark halo white dwarfs by
searching for nearby counterparts in existing, ground-based proper motion
surveys. We find no candidate counterparts, although the volume probed by these
surveys appears to be significantly larger than that probed by the Hubble Deep
Field.

  The power of a survey to detect objects of absolute magnitude $M$ visible to
limiting magnitude $m$, assumed to have a constant density is the effective
volume of the survey,

\begin{equation}
\veff= \frac{\Omega}{3} 10^{0.6(m-M)+3} \epsilon ~{\rm pc^3}.
\end{equation}

where $\epsilon$ is the efficiency (completeness) of the survey and $\Omega$ its
solid angle in steradians.  

  We examine two photographic surveys of proper motions, the Luyten Half Second
Catalogue (1979, LHS) and the Knox, Hawkins \& Hambly (1999, KHH).  We show
that both surveys have greater power to detect intrinsically faint high proper
motion objects than the IRGS99 survey. 

  In neither ground-based survey do we find candidates for local counterparts to
the proposed dark halo WDs in HDF.  The probability of finding objects in the
less powerful HDF and not in the ground-based surveys is low, suggesting that
the HDF objects are not dark halo white dwarfs.

  This paper is organised as follows. In section 2 we discuss how to locate
nearby counterparts to the proposed white dwarfs in proper motion surveys. In
section 3 we search two such surveys, but find no candidates for such objects,
while also demonstrating that both surveys would be more efficient at locating
them than HDF itself.  In section 4 we discuss our results in terms of models
of cooling white dwarfs, and in section 5 we draw our conclusions.

\section{Proper motion search for dark halo white dwarfs}

 Several authors have already searched for possible nearby dark halo objects,
but have found no candidates, allowing an upper limit to be placed on their
luminosity. The Luyten Half Second Catalog (LHS) has been analysed in this
manner by Graff, Laughlin and Freese (1997), Fuchs and Jahrei\ss\ (1998) and
Hansen (1999b).  We follow here a similar approach as these authors while
adapting it to the particular details of the IRGS99 proposal.

  Assuming that some fraction $\fwd$ of the dark halo is composed entirely of
white dwarfs of mean mass $M_{\rm WD}$, and that $F_H$ of these dwarfs have
hydrogen atmospheres, then the number $N_{\rm WD}$ of WDs we expect in the
survey is:

\begin{equation}
N_{\rm WD} = \frac{\rho_H}{M_{\rm WD}}\veff \fwd F_H.
\end{equation}

where $\rho_H$ is the local halo dark matter density.  The dark halo is here
assumed to have constant density, an adequate approximation out even to the
range of the HDF, a kiloparsec from the Sun.

  In two of the proper motion surveys discussed in the next section, there is
an upper limit on the detectible proper motion $\mu_{\mathrm max}$. In order to
estimate the number of WDs with proper motions $\mu < \mu_{\mathrm max}$, we
assume the dark halo WD system is an isothermal sphere and that the 1-D
velocity dispersion is $220/\sqrt{2} = 156$ \kms.  We assume the system is
non-rotating, has local density $\rho_H = 0.0076$ M$_\odot$ pc$^{-3}$, and that
the mean WD mass is $M_{\mathrm WD} = 0.66$ M$_\odot$. We then determine the
fraction of stars with proper motion $\mu < \mu_{\mathrm max}$ from the
velocity distribution, while accounting for the Solar motion (i.e. we use Eq. 6
of Fuchs and Jahrei\ss\ 1998).

  We search for nearby counterparts to the proposed HDF white dwarfs using the
reduced proper motion, $H$ (Luyten 1922, Evans 1992), which is the proper
motion equivalent of absolute magnitude.  For a star of absolute magnitude $M$
and transverse velocity $V_T$ (in \kms), or apparent magnitude $m$ and proper
motion $\mu$ (in arcsec/year), $H$ is

\begin{equation}
H = M + 5 {\mathrm log} V_T - 3.379 = m + 5 {\mathrm log} \mu + 5.
\end{equation}

  The reduced proper motions of the objects detected by IRGS99 in the HDF lie
in the range $24 < H_R< 26.5$, as expected for objects with velocities
characteristic of the dark halo and absolute magnitude at $M_V \approx 17.5$,
(typical of the WDs proposed by IRGS99).  Nearby counterparts to the HDF
objects would also have reduced proper motions in this range.

\section{The three proper motion surveys}

  We describe three surveys in which one can search for dark halo white dwarfs,
two of which are ground-based and would locate nearby objects, and the third
being the HDF itself (which has been searched by IRGS99). For each survey we
calculate the effective volume probed. For the two ground-based surveys we
search for but locate no dark halo white dwarf candidates.

\subsection{The Hubble Deep Field}

  The Hubble Deep Field is the deepest search for any object, and covers a
comparatively small solid angle, only 4.4 arcmin$^2$, or $5\times 10^{-8}$ of
the angle covered by the larger of the two ground-based surveys (LHS).  IRGS99
effectively run two experiments in searching for faint moving objects in HDF.
They are most confident of their results for $I<28$, and find two objects with
$I<28$.  One of the objects varies in magnitude, and has the wrong $B-V$
colours to be a white dwarf, leaving one good candidate object, 4-551.
Extension of the survey out to $I<29$ reveals three additional candidates,
though only one of these has a secure proper motion.  The completeness of the
Ibata \etal survey is $42 \pm 2 \%$ for objects in the range $27<I<28$. The
effective volume probed by HDF (using eqn 1) for objects at $M_V = 17.5$ is
$v_{\mathrm eff}^{\mathrm HDF} = 225 \times 0.42 = 95 $ pc$^3$. Conservatively,
no correction has been included for the proper motion window of the HDF survey
even though the measured proper motions are only $\sim 2$ times larger than the
minimum proper motion. IGRS99 do not discuss the upper proper motion limit of
their survey, but it is likely to be much larger than the proper motions of the
WDs.

\subsection{ESO/SERC Area 287}

  Knox, Hawkins and Hambly (1999, hereafter KHH) have recently surveyed
``ESO/SERC Area 287'', searching for faint high proper motion objects in order
to locate the end of the disk white dwarf cooling sequence. They used about 100
UK Schmidt $R$-band exposures taken at a range of different epochs, which they
stacked in three different manners either to go as deep as possible $(R=22)$ or
to be able to recover stars with high proper motions (10 arcsec/yr).

  The three KHH proper motion experiments have different apparent magnitude
$R$-band limits and upper proper motion limit $\mu_{\mathrm max}$, and are
denoted by (i), (ii) and (iii). Hambly (1999, private communication), has
undertaken a careful re-analysis of the original data as a result of the IRGS99
results, and has advised us that the effective area of the survey should be
conservatively set at 12 square degrees. The three experiments are summarised
in Table 1, where we show the upper limit on proper motion $\mu_{\mathrm max}$,
the $R$-band apparent magnitude limit and effective volumes probed for each
survey (including the effect of the proper motion window) $v_{\mathrm eff}$.
The most effective of the surveys is (i), for which $v_{\mathrm eff}^{\mathrm
KHH} = 198 $ pc$^3$ for white dwarfs at $M_R = 17.5$.

\begin{table} 
\caption{Three proper motion surveys of ESO/SERC field 287 from Knox et al
(1999), showing the upper proper motion limit $\mu_{\mathrm max}$, apparent
$R$-band magnitude limit, fraction of sources expected in the proper motion
window $\epsilon$ (i.e. with $\mu < \mu_{\mathrm max})$ and the effective
volume $\veff$ for WDs at $M_R=17.5$}
\begin{tabular}{l|cccc}
ID & $\mu_{\mathrm max}$ & $R_{\mathrm lim}$ & $\epsilon$ & $\veff$   \\
   &     arcsec/yr       &                   &            &  pc$^3$  \\
\hline
(i)     &  10   & 21.2 &1.00 & 198 \\
(ii)    &  1.9  & 21.2 &0.83 & 168 \\
(iii)   &  0.5  & 22.0 &0.24 & 146 \\
\hline
\end{tabular}
\end{table}

  KHH show reduced proper motion in the $R$-band, $H_R$ versus colour diagrams
for their sources (their figures 9, 10 and 11). Inspection of their plots shows
there are no sources with $H_R > 24$, (blue or otherwise).  Almost all their
sources have $H_R < 20$ (as expected for late type dwarfs and disk white
dwarfs). We conclude there are no candidate dark halo WDs in KHH.

\subsection{Luyten Half Second Survey}

  The Luyten Half Second catalog (LHS --- Luyten 1973) is a proper motion
survey of most of the sky, complete in the range $0.5 < \mu < 2.5$, which was
obtained by blinking Palomar plate pairs.  Dawson (1986) has studied in detail
the completeness of the LHS, finding that it is 90 \% complete to Luyten
$R$-band magnitude $R_L = 18$ (for declination $\delta > -30^\circ$, and
Galactic latitude $|b| > 10^\circ$).

Salim and Gould (1999) have recently studied the completeness of the NLTT
(``New Luyten Catalog of stars with proper motions larger than Two Tenths of an
arcsecond'') in order to estimate the self lensing rate of field stars for
astrometric microlensing. The NLTT is the extension to lower proper motions of
the LHS catalog, and we have used the NLTT to determine the completeness of
LHS. We find that LHS is 60\% complete down to Luyten $R$-band magnitude $R_L =
18.5$.  In Appendix A, we give a detailed presentaion of this determination.
Furthermore, in Appendix B, we calibrate the Luyten $R$-band magnitude, $R_L$,
finding that it is actually closer to Johnson $V_J$ than to Cousins $R_C$. The
calibration is

\begin{equation}
R_L = V_J - 0.37(V_J-R_C) + 0.06.
\end{equation}

  This relation allows us to compare the LHS survey directly with HDF's
$V$-band magnitude (and is particularly useful since HDF was imaged in $U$,
$B$, $V$ and $I$ but not $R$).

  We obtained the LHS catalog from the SIMBAD data center, and show in figure 1
the reduced proper motion $H$, computed in Luyten's $R$-band magnitude $R_L$
versus his $m_{\mathrm pg}-m_R$ colour index. The disk main sequence (running
from upper-left to lower-right) and the disk white dwarf cooling sequence
(lower-left) are seen in the figure. There are no sources with $H > 24$.
Dark halo white dwarfs are expected to have $H \ga 25$.  We note that there
are a few objects in the LHS with $H > 23.5$.  A literature search using
SIMBAD showed that all could be firmly identifications as either M dwarfs or
disk white dwarfs through parallax, photometry and/or spectroscopic methods
(Bessell, 1991, McCook and Sion 1987).  We conclude that there are no
counterparts in the LHS to the moving sources in the HDF, confirming previous
studies (Graff, Laughlin and Freese 1997, Fuchs and Jahrei\ss\ 1998, Hansen
1999b).

  Richer et al (1999) discuss the colours of the white dwarfs proposed by
IRGS99.  In the age range of interest, 7 to 15 Gyr, the models have $V-R$
colours in the range $0 < V-R < 0.5$. We discuss in much more detail the
colours of the models in the next section, but at this point if we
conservatively adopt $V-R=0$, then the $V$-band limit of LHS is (from Eqn 4) is
$V = 18.4$. Taking this magnitude limit, accounting for the fraction of sky
covered, $\sim 8.5$ sr, and the completeness (60\%), the Luyten catalogue
probes an effective volume of $v_{\mathrm eff}^{\mathrm LHS} = 5700$ pc$^3$ for
$M_V = 17.5$ dark halo white dwarfs.  Applying the proper motion window for
sources in LHS $(0.5 < \mu < 2.5)$ we obtain an survey efficiency of $\epsilon
= 0.22$, which reduces the effective volume to

\begin{equation}
\veff^{\rm LHS} = 1290 {\rm pc}^3
\end{equation}

which is much greater than the effective volume of both the KHH survey (198
pc$^3$), and the HDF survey (95 pc$^3$).

  We conclude that the type of objects seen by IRGS99 should be present in
significant numbers in the LHS catalog with $H_R \ga 25$. The lack of such
objects in LHS argues against the interpretation that they are dark halo white
dwarfs.

\begin{figure}
\epsfig{file=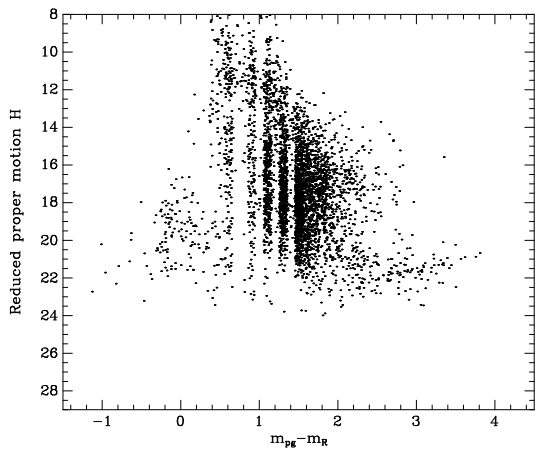,width=70mm}
\caption{The reduced proper motions $H$ in the Luyten $R$-band for the stars in
the Luyten Half Second survey versus their colour index $(m_{\mathrm
pg}-m_R)$. No sources are found with $H > 24$, where dark halo white dwarfs are
expected to lie. Note that the verticle stripes are an artifact due to the
limited colour resolution of the LHS survey. A small random number has been added
to the Luyten colours to reduce crowding in the figure}
\end{figure}

\section{Discussion}

  The three surveys are summarised in Table 2.  The combined ground-based
surveys probe an effective volume which is larger than that probed by HDF by a
factor of

\begin{equation}
\label{ratio}
18 \times 10^{0.6 (R_L-I)} \frac{\epsilon_{\rm LHS}}{\epsilon_{\rm HDF}}+ 1.7
\times 10^{0.6 (R-I)} \frac{\epsilon_{\rm KHH}}{\epsilon_{\rm HDF}}
\end{equation}

  where $R_L$ is the Luyten $R$-band magnitude, $\epsilon$ is the survey
efficiency, which includes the completeness and the proper motion window, and
$R-I$ is the WD colour.

\begin{table} 
\caption{Summary of the limits of the three experiments discussed in the Paper.
The efficiency includes the survey completeness and the probability that an
object at absolute magnitude 17.5 will have a proper motion within the proper
motion window of the survey}
\begin{tabular}{l|ccc}
          & $\Omega$            & Mag.     & Efficiency   \\
Survey    &  str                & limit    & $\epsilon$   \\
\hline
HDF       & $3.7\times 10^{-7}$ & $I<28$   & 0.42         \\
KHH (i)   & $7.6\times 10^{-3}$ & $R<21.2$ & 1.00         \\
LHS       & 8.5                   & $R_L<18.5$ & 0.13       \\
\hline
\end{tabular}
\end{table}

  As can be seen from Eq. 6, the two important parameters in determining the
relative strengths of the ground-based surveys to the HDF are the efficiency,
$\epsilon$ and the colour of the dwarfs.  We next discuss the effect of the
colours of the WDs in detail.

\subsection{White Dwarf models}

  IRGS99 propose that the faint blue sources which they have found in the HDF
might be old, hydrogen-atmosphere white dwarfs, which have the surprising
9property of being blue (Hansen 1999b) due to $H_2$ opacity (old
helium-atmosphere white dwarfs would have cooled so effectively that they would
not be visible in any existing survey).  Hansen (1999b) has computed $V$- and
$I$-band absolute magnitudes for old white dwarfs for a range of ages. Over the
age range of interest, 11 to 16 Gyr, the $V$-band absolute magnitude of
Hydrogen atmosphere white dwarfs is in the range $17 \la M_V \la 18$ while the
colours lie in the range $-1 \la V-I \la 1$. These properties are consistent
with the interpretation of the moving HDF objects as old dark halo white
dwarfs.

  We adopt the models of Hansen (1999b), who kindly made unpublished $B$ band
colours available to us.  Other observables of these models are discussed in
Richer et al (1999).  Within these models, once the temperature of the white
dwarf cools below 4000K, the spectral energy distribution becomes extremely
non-blackbody.  Most of the light is emitted in the $V$ and $R$ bands with the
peak shifting to the {\it blue} as the star cools, and the absolute $M_V$
magnitude of the star stays roughly constant.  Thus, the effective volume
probed by the photographic catalogues does not strongly depend on models, while
the IRGS99 volume, which has an $I$ band magnitude limit, depends strongly on
the temperature of the white dwarf.  When computing the relative strengths of
the different surveys, the most important parameter is the $V-I$ colour.

  We will discuss three models which cover the $V-I$ colours of cool white
dwarfs.  We examined several other models with different ages and white dwarf
masses, the three models we discuss illustrate the reasonable parameter space
since only the $V-I$ colour plays a significant role.  The three models are
denoted O, R and B, where model O, is a fit to the observed candidate
white dwarf 4-551, model R represents a dwarf which is red in $V-I$, and
model B, a dwarf which is blue in $V-I$.  The parameters chosen for these
dwarfs are shown in Table 3.

  Note that model R is the reddest possible hydrogen atmosphere dwarf.  Both
cooler and hotter dwarfs are bluer.

  Calculations of effective volumes for different models and different surveys
are shown in Table 4.  In all cases, the LHS catalogue is the most potent
survey, then the KHH survey, and the IGRS survey is the least potent.  The
combined photographic surveys are $7 - 18$ times as powerful as IRGS99 to
$I=28$.

\subsection{Combining the surveys}

  Na\"\i vely, if the object 4-551 detected by IRGS99 is typical of the halo
population, there should be tens of dwarfs in the ground-based surveys.
Instead, there are none.  We calculate the probability of such a mismatch
between surveys as follows:

  Let $\lambda$ be the mean expected number of dwarfs seen by IRGS99.  We define
$\alpha$ to be the ratio of effective volumes probed by the different surveys:

\begin{equation}
\alpha=\frac{{\veff^{\rm LHS}} + \veff^{\mathrm KHH}}{\veff^{\mathrm HDF}}
\end{equation}

so that the mean expected number of dwarfs in the combined photographic surveys
is $\alpha\lambda$.  Then the probability that at least one dwarf will be seen in
the HDF is $P_{\rm HDF} = 1 - e^{-\lambda}$, while the probability that no
dwarfs will be seen in either photographic surveys is $P_{\rm phot} =
e^{-\alpha\lambda}$.

The combined probability of both events is

\begin{equation}
P=P_{\rm HDF} P_{\rm phot} = (1 - e^{-\lambda})e^{-\alpha\lambda}.
\end{equation}

This probability is maximized when 

\begin{equation}
\label{lambdamax}
\lambda_{\rm max} = \ln{\frac{\alpha+1}{\alpha}}
\end{equation}

and has a value of

\begin{equation}
P_{\rm max} = \frac{\alpha^\alpha}{(\alpha+1)^{\alpha+1}} \sim
\frac{1}{e(\alpha + 1)} \, .
\end{equation}

  Note that, as shown in Table 4, even for the reddest model, where the HDF is
relatively most effective, the probability of seeing a star in the HDF and no
stars in the more powerful photographic surveys is only 5\%.  In the case of
the actual star observed, 4-551, the probability is lower (2\%) that no other
stars of the same type would be observed in the photographic surveys.

\begin{table} 
\caption{Three models of halo white dwarfs from Richer \etal (1999). We
consider a red (``R'') and a blue (``B'') model, and a model which is a match
to one of the observed (``O'') objects in the HDF (object 4-551). }
\begin{tabular}{lrrr}
\hline
                   & \multicolumn{3}{c}{White Dwarf Model} \\
Parameter          &  R & O & B \\
\hline
Mass ($M_\odot$) &  0.66 &  0.70 &  0.80  \\
Absolute Magnitude $M_V$     &  17.49&  17.40&  18.01  \\
Age (Gyr) &  12.0 &  12.1 &  12.0  \\
$V-I$     &  1.10 &  0.40 &  0.02  \\
$V-R$     &  0.96 &  0.57 &  0.73  \\
\hline
\end{tabular}
\end{table}

\begin{table} 
\caption{The first three rows show the effective volume of the three surveys to
the three WD types considered in Table 3. The effective volume includes the
survey completeness and the reduction due to the proper motion window.  Row 4
shows the combined effective volume of the surveys. Row 5 shows $N_{\rm WD}$,
the number of Hydrogen atmosphere white dwarfs expected in the combined surveys
(see section 4.3) assuming they make up 50\% of the dark halo density.  Row 6
shows the ratio of effective volume for the combined ground-based surveys to
the HDF survey: they are typically 10-20 times more powerful than the HDF
survey.  Row 7 shows the (maximised) probability that objects would be detected
in the HDF but not detected in the ground-based surveys, and is typically below
5\%. The probability has been maximised by fitting for the dark halo mass
fraction in white dwarfs, which is shown in row 8.}
\begin{tabular}{l|ccc}
\hline
                       & & WD Model & \\
Description            & R & O & B \\
\hline
$v_{\rm eff}^{\rm HDF}$ (pc$^3$)   &  438   &      189 &         48  \\
$v_{\rm eff}^{\rm LHS}$ (pc$^3$)   & 2730   &     3220 &        832  \\
$v_{\rm eff}^{\rm KHH}$ (pc$^3$)   &  403   &      121 &         52  \\
Total Volume (pc$^3$)              & 3571   &     3530 &        932  \\
$N_{\rm WD}$                       &   27   &       26 &          7  \\
\hline
(LHS + KHH)/HDF&      7.2&     17.7&     18.4 \\
\hline
Probability    &     5.0\%&     2.0\%&     2.0\% \\
Halo Fraction  &     2.0\%&     2.0\%&     3.0\% \\
\hline
\end{tabular}
\end{table}

\subsection{Halo Fraction}

  Having calculated the effective volumes of the surveys, we can calculate the
number of dwarfs that would be visible assuming the halo were composed of white
dwarfs. This assumption is not entirely consistent with the microlensing
results --- the MACHO microlensing experiments curently suggest that half the
dark halo could be in the form of $\approx 0.5 \Msun$ mass objects (Alcock
et.al 1997), whereas EROS2 (Afonso et.al, 1999) concludes that objects of this
mass can be ruled out at the 95\% confidence level. We do not constrain the
calculations in this section by these results, but rather use the results of
the HDF proper motion search itself as our starting point. We adopt a local
halo density of $0.0076 M_\odot\, {\rm pc}^{-3}$ and assume that 50\% of this
density is due to Hydrogen atmosphere white dwarfs.  We assume that the
remaining 50\% of the dark halo is in helium atmosphere white dwarfs, which
will have cooled far below the detection limits of the surveys. 

The number of hydrogen atmosphere white dwarfs expected in the combined surveys
$N_{\rm WD}$ is shown in row 5 of Table 4. The probability that one or more
white dwarfs would be seen in HDF while none are seen in the photographic
surveys is very low for all the models. Using equation 8, all three models in
which Hydrogen atmosphere white dwarfs make up half of the dark halo can be
ruled out with greater than 99\% confidence.

We tested models in which the dark halo white dwarf fraction maximises the
probability that one or more white dwarfs would be seen in HDF while none are
seen in the photographic surveys (Eqns 9 and 10). Model R has the highest
probability of explaining the combined survey results, albeit with a low
probability of only 5\% and a dark halo white dwarf fraction of just 2.0\%. For
the other two models the probability of there being at least one object in HDF
and none in the ground based surveys is $P_{\rm max} = 2\%$, and the
corresponding fraction of dark hao white dwarfs is also very low, less than
3\%.

\subsection{Survey in progress: EROS-II Wide Field Imager}

  A $V$-band limit on the luminosity of putative dark halo white dwarfs has
been set by Goldman (1999), using the first 140 square degrees of the EROS-II
survey (a $V$ and $I$ band wide field imager), which will cover 350 square
degrees and reach $I \approx 20.5$ and $V \approx 21.5$ when completed. No high
proper motion objects were detected, and Goldman (1999) uses this to set a
$V$-band absolute magnitude limit of $M_V > 17.2$ on dark halo WDs. This is
consistent with the Ibata et al proposed WDs, since they lie in $17 \la M_V \la
18$. The survey is currently approaching completion and will be very sensitive
to nearby white dwarfs, surveying some 5000 pc$^3$ (Goldman, private
communication).

\section{Conclusions}

  Ibata et al (1999) have recently discovered faint moving objects in the
Hubble Deep Field, proposing that these might be cool white dwarfs making up
the entire mass of the Galactic dark halo. We have searched for nearby
counterparts to these objects in a number of ground-based proper motion
surveys. No such objects have been found, even though the combined photographic
surveys are tens of times more powerful than the HDF.  The probability of this
occuring is quite low, $<5\%,$ even in the most conservative model.  This study
leads us to the conclusion that it is unlikely that hydrogen atmosphere white
dwarfs make up a significant fraction of the halo dark matter.  No limits can
be placed yet on helium atmosphere dwarfs from optical searches.

\section*{Acknowledgments}

  We thank Bertrand Goldman, Andy Gould, Geza Gyuk, Nigel Hambly, Hartmut
Jahrei\ss, Harvey Richer and Ali Talib for help and interesting discussions.
Brad Hansen shared his unpublished calculations of $B$-band colours of cool
white dwarf atmospheres.  This research was supported in part by the Academy of
Finland, and by Danmarks Grundforskningsfond through its support for the
establishment of the Theoretical Astrophysics Center. We have made extensive
use of the SIMBAD astronomical data base for which we are very grateful.

\appendix
\section{Completeness of NLTT}

In this section we perform a statistical test to investigate the completeness
of the faint end of NLTT down to its nominal cutoff of $\mu_1= 200\, \masyr$.
In the test we assume that the local luminosity function is constant, and that
the number density of stars does not change appreciably on scales equivalent to
a distance modulus of 0.5 mag.

Consider two spheres centered around the Sun, the volumes of which stand in
ratio 2:1. This is equivalent to radii being in relation $r_1/r_2=1.259$, or
distance modulus difference of 0.5 mag. If we define the outer edge of the
bigger sphere as the distance at which a star of apparent magnitude $R_{\rm
L,1}$ produce a proper motion $\mu_1= 200\, \masyr$, then this same star, if
placed at distance $r_2$, would have a proper motion of $\mu_2 = {r_1 \over r_2
} \mu_1 = 252\, \masyr$. Also, it would be 0.5 mag brighter. Therefore, $\mu_2$
defines a proper motion limit at the distance $r_2$ that is equivalent to
proper motion limit $\mu_1$ at $r_1$. These are the lower limits. However, in
NLTT there is also an upper proper-motion cutoff of $\mu_2^{\rm lim}= 2500\,
\masyr$.  If we adopt this as a limit, this corresponds to some inner boundary
of the smaller sphere (which we can now call a shell). Everything closer than
this inner boundary would have $\mu>\mu_2^{\rm lim}$ and would not be included
in NLTT. Now, in order to keep volumes of both shells in appropriate ratio, the
outer sphere (shell) has to have an inner edge corresponding to a proper motion
of $\mu_1^{\rm lim} = {r_2 \over r_1 }\mu_2^{\rm lim} = 1986\, \masyr$.

Now that we have defined the two shells in terms of the limiting proper
motions, the statistical test consists of comparing the number of stars $N_1$
of a given magnitude $R_{\rm L}$ (in a $\Delta R_{\rm L}=0.5\,$ mag bin) in the
outer shell ($200\,\masyr < \mu < 1986\,\masyr$), with the number of stars
$N_2$ of a magnitude $R_{\rm L}-(R_{\rm L, 1}-R_{\rm L, 2}) = R_{\rm L}-\Delta
R_{\rm L} = R_{\rm L}-0.5$ in the inner shell ($252\,\masyr < \mu <
2500\,\masyr$). The 0.5 mag shift (equal to one bin) brings the absolute
magnitudes of stars in the outer shell to that of the inner shell. The measure
of completeness at magnitude $R_{\rm L}$ is given by the ratio

\be 
f(R_{\rm L}) = {N_1(R_{\rm L}) \over N_2(R_{\rm L}-0.5)}.
\label{eqn:c1}
\ee

\begin{figure}
\epsfig{file=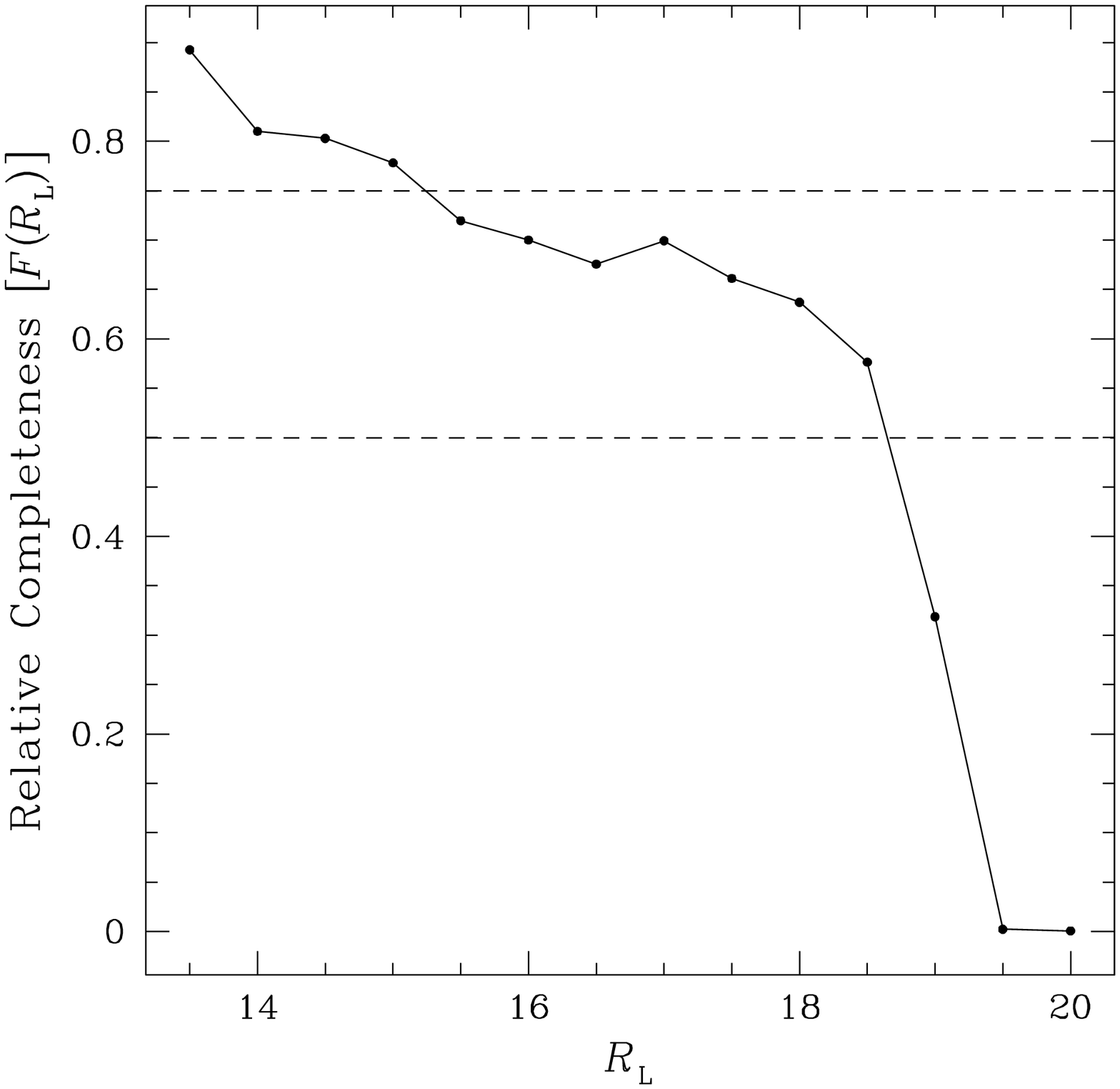,width=70mm}
\caption[junk]{\label{fig:A1} Completeness of NLTT relative to $R_{\rm L}=13$,
in the Completed Palomar Region ($\delta \ga -33^{\circ}$, $\mid
b\mid>10^{\circ}$). Dashed lines show 75\% and 50\% completeness levels.  }
\end{figure}

If the sample of stars of apparent magnitude $R_{\rm L}$ is 100\% complete with
respect to those of $R_{\rm L}-0.5$, then $f(R_{\rm L})\equiv (r_1/r_2)^3 =
2$. Now we can define the completeness function $F(R_{\rm L})$ for the stars of
apparent magnitude $R_{\rm L}$, in the following way

\be
F(R_{\rm L}) = \prod^{R_{\rm L}'=R_{\rm L}}_{R_{\rm L}'=R_{\rm L, comp}+\Delta
R_{\rm L}}{f(R_{\rm L}') \over 2},
\label{eqn:c2}
\ee

where $R_{\rm L, comp}$ is some bright apparent magnitude at which we believe
the catalogue is complete.

In Figure \ref{fig:A1} we show the completeness function $F(R_{\rm L})$ for the
faint end of NLTT. More specifically, the test was performed on the subsample
of NLTT that is believed to be spatially complete, that is, the part called the
Completed Palomar Region (CPR) by Dawson (1986). This region covers northern
declinations ($\delta \ga -33^{\circ}$), and avoids the galactic plane ($\mid
b\mid>10^{\circ}$). We take $R_{\rm L, comp}=13$. The choice is somewhat
arbitrary, but we have reasons to believe that NLTT is complete at this
magnitude. First, when we plot $f(R_{\rm L})$ against $R_{\rm L}$, we get a
flat region around $R_{\rm L}=13$. Going to still brighter magnitudes might
bring us into the part of NLTT that was not compiled from the photographic
plates. Therefore, Figure \ref{fig:A1} shows the completeness of $R_{\rm L}$
with respect to $R_{\rm L}=13$. Dashed lines represent 75\% and 50\%
completeness levels. The completeness drops gradually from 90\% at $R_{\rm
L}=13.5$ to 60\% at $R_{\rm L}=18.5$. 

\section{Photometric calibration of NLTT}

\begin{figure}
\epsfig{file=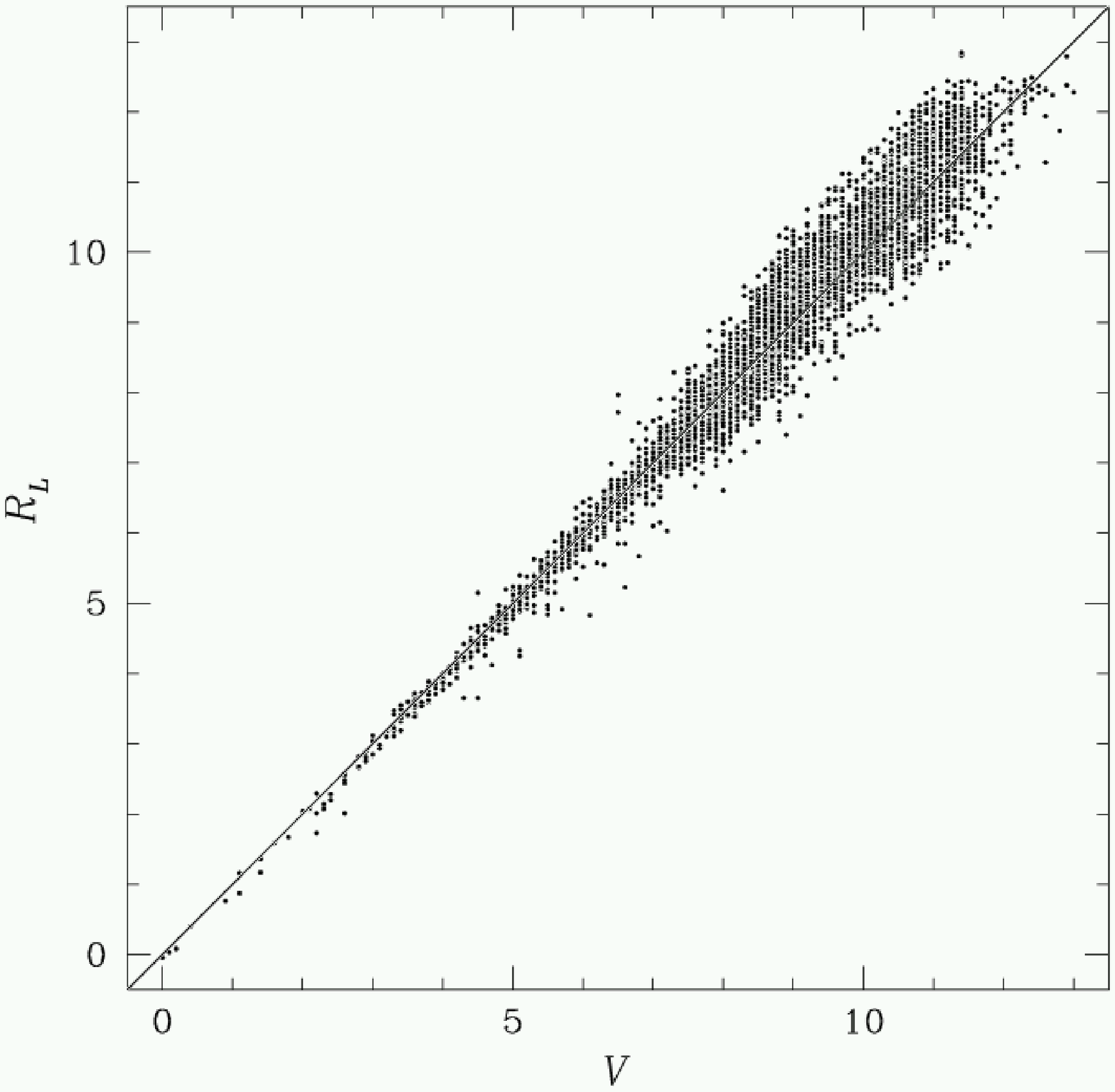,width=70mm}
\caption[junk]{\label{fig:A2} Luyten's ``red'' magnitude $R_L$ versus Johnson
$V$ magnitude for stars in both NLTT and Hipparcos catalogs. A 1:1 line is
shown for reference. }
\end{figure}

\begin{figure}
\epsfig{file=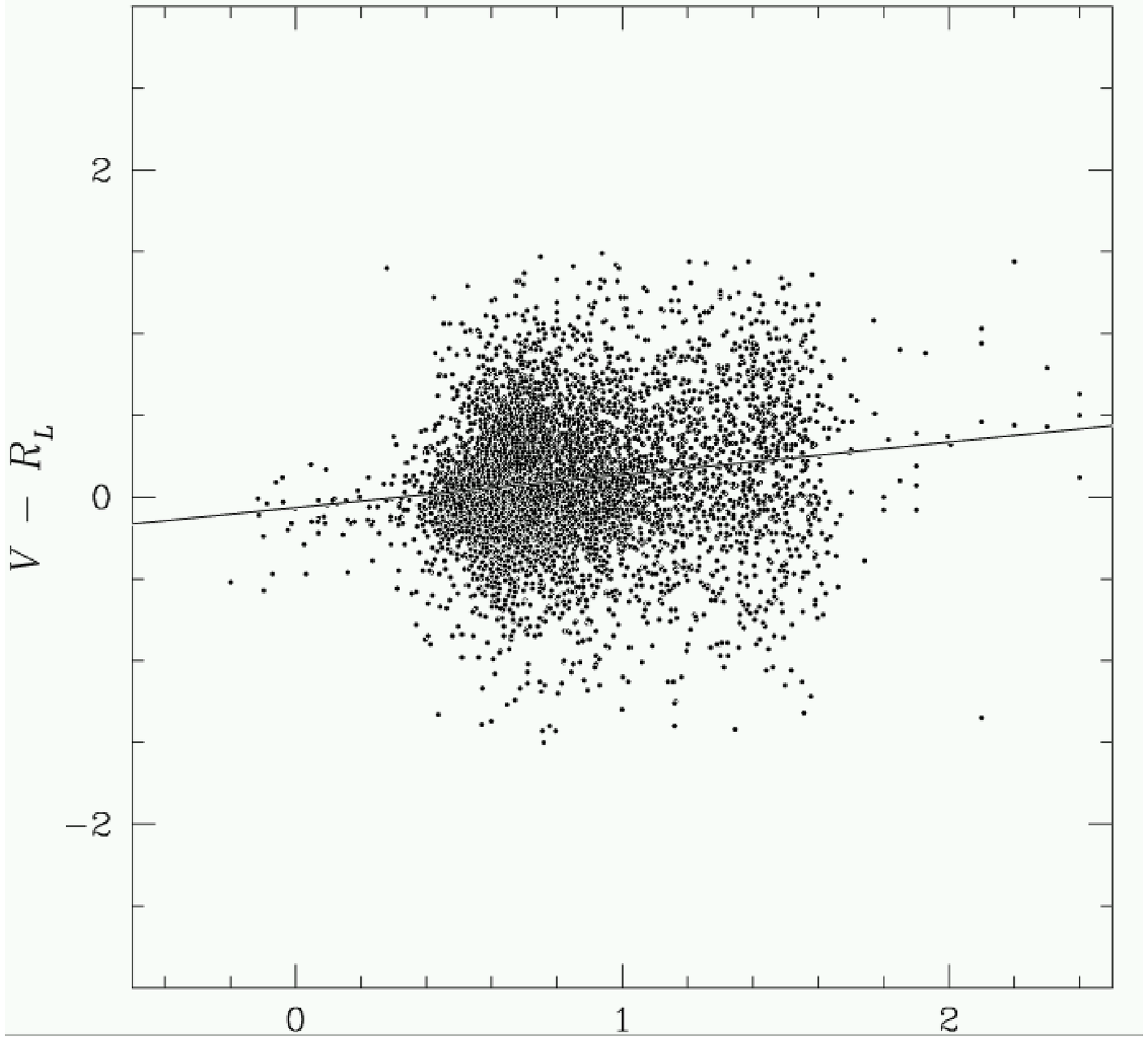,width=70mm}
\caption[junk]{\label{fig:A2} Residuals between Luyten's $R_L$ magnitude and
Johnson $V$ magnitude plotted against the Johnson $B-V$ colors of the stars
that appear in both NLTT and Hipparcos catalogs. The best-fit line is the
calibration given in Equation \ref{eqn:hip1}.  }
\end{figure}

Throughout the previous section we used Luyten's red magnitude $R_{\rm L}$. We
derive here a calibration of $R_{\rm L}$ to standard Johnson magnitudes.

NLTT magnitudes are given as photographic (blue plate) and red plate
magnitudes. The Hipparcos catalog contains most of the NLTT stars to its
detection limit ($V\sim 12$). We matched NLTT stars with the corresponding
Hipparcos stars (details are given in Salim \& Gould 1999), and found 6084
matches with the complete photometric information. These stars therefore
calibrate the bright end of NLTT ($0<V<12.5$), as follows:

\be
V = R_{\rm L} - 0.06 + 0.200 (B-V)
\label{eqn:hip1}
\ee

and 

\be V = R_{\rm L} - 0.08 + 0.196 (B_{\rm L}-R_{\rm L}),
\label{eqn:hip2}
\ee

where $B_{\rm L}$ and $R_{\rm L}$ are NLTT's blue (photographic) and red
magnitudes, respectively. The first relation is shown graphically in Figure
\ref{fig:A2}. From the first relation we can see that both Johnson $V$ and
Luyten's $ R_{\rm L}$ have almost the same zero points. More importantly, the
low color term of 0.2 puts $R_{\rm L}$ magnitudes much closer to $V$ than to
standard Kron $R$. This is in sharp contrast to Dawson (1986) who finds that
$R_{\rm L}$ and Kron $R$ are almost the same, with the only difference being in
zero point. Dawson's calibration would give a colour term coefficient of about
0.6 in eq.\ \ref{eqn:hip1}, not 0.2 that we obtain. RMS in both equations
\ref{eqn:hip1} and \ref{eqn:hip2} are 0.40 mag. We cannot account for this
discrepency, but note that it would not change the main conclusions of the
paper.

The calibration above is restricted to the bright end of NLTT. Obtaining a
calibration for the fainter part was somewhat more complicated.  Our faint end
calibration is based on the USNO-A2.0 all-sky astrometric survey (Monet
1998). Although USNO-A2.0 itself does not contain standard magnitudes, it can
be calibrated independently (see Salim \& Gould 1999). The USNO-A2.0 catalog's
photometric accuracy ($\approx 0.25$ mag) is far superior to NLTT's photometry
($\ga 0.5$ mag). The complicated step is finding faint NLTT stars in the
USNO-A2.0 catalog. This procedure is also described in Salim \& Gould
(1999). In the end, we identify 33286 NLTT stars (in the northern,
($\delta>-15^{\circ}$) part of the sky) in USNO-A2.0. Out of this number, 21053
stars have $V>15$ and we use them to construct the calibration. We obtain the
following relation

\be
V =  R_{\rm L} + 0.01 + 0.230 (B-V)
\label{eqn:usno}
\ee

with an RMS scatter of 0.47 mag. This confirms the calibration obtained at the
bright end, and reaffirms our suggestion that $ R_{\rm L}$ is closer to $V$
than to Kron $R$. We use calibration from eq.\ \ref{eqn:hip1} throughout this
paper.


\begin{thebibliography}{}

\bibitem[]{} Afonso,  et.al. (The EROS Collaboration), 1999, A\&A, 344, L66 

\bibitem[]{} Alcock C. et al. 1997, ApJ, 486, 697

\bibitem[]{} Bessell, M. S. 1991, AJ, 101, 662

\bibitem[]{} Chabrier, G. 1999, ApJ, 513, 103

\bibitem[]{} Conti, A., Kennefick, J. D., Martini, P. \& Osmer, P.S. 1999, AJ,
117, 645

\bibitem[]{} Dawson, P. C. 1986, ApJ, 311, 984

\bibitem[]{} Evans, D.~W., 1992, MNRAS, 255, 521

\bibitem{} Fields, B., Freese, K. \& Graff, D.S., 1998, New Astro., 3, 347

\bibitem{} Fields, B., Freese, K. \& Graff, D.S., 1999, ApJ, in press

\bibitem[]{} Fuchs, B. and Jahrei\ss\, H. 1998, A\&A 329, 81

\bibitem[]{} Gliese, W. and Jahrei\ss\, H. 1980, A\&A 85, 350

\bibitem[]{} Goldman, B. 1999, 3rd Stromlo Symposium, The Galactic Halo, ASP
Conference Series, Vol 165, p413, B.~Gibson, T.~Axelrod, M.~Putman Eds.

\bibitem[]{} Graff, D.S., Laughlin, G., \& Freese, K., 1998, ApJ, 499, 7

\bibitem[]{} Graff, D.S., Walker, T.P., Freese, K. \& Pinnsoneault, M.H., 1999,
ApJ, 523, 77

\bibitem[]{} Hambly, N. C., Smartt, S.  J., Hodgkin, S. T., Jameson, R. F.,
Kemp, S. N., Rolleston, W. R. J. \& Steele, I. A. 1999, MNRAS, 309, L33

\bibitem{} Hansen, B. 1999a, ApJ, 517, 39

\bibitem{} Hansen, B. 1999b, ApJ, 520, 680

\bibitem[]{} Harris, H. C., Dahn, C.  C., Vrba, F. J., Henden, A. A., Liebert,
J. , Schmidt, G. D. \& Reid, I. N.  1999, ApJ, 524, 1000

\bibitem[]{} Hodgkin, S.T., Oppenheimer, B.R., Hambly, N.C., Jameson, R.F.,
Smartt, S.J., Steele, I.A. 1999, preprint

\bibitem[]{} Ibata, R., Richer, H., Gilliand, R. \& Scott, D. 1999,
astro-ph/9908270 (IRGS99)

\bibitem[]{} Knox, R., Hawkins, M. \& Hambley, N. 1999, MNRAS, 306, 736 (KHH)

\bibitem[]{} Luyten, W. \& La Bonte, A., 1973, {\it The South Galactic Pole},
Univ. of Minnesota, Minneapolis

\bibitem[]{} Luyten, W., 1922, {\it Lick Obs Bull}, No. 336

\bibitem[]{}Monet, D.\ 1998, BAAS, 193, 120.03

\bibitem[]{} McCook, G. P. \& Sion, E. M. 1987, ApJSupp, 65, 603

\bibitem[]{} Richer, H.B., 1999, astro-ph/9906424

\bibitem[]{} Richer, H.B., Hansen, B., Limongi, M., Chieffi, A,, Straniero,
O. \& Fahlman, G. G. 1999, ApJ, in press, astro-ph/9908337

\bibitem[]{}Salim, S. \& Gould. A.\ 1999, astro-ph/9909455


\end{thebibliography}
\end{document}